\documentclass[10pt]{npqcd}
\begin{document}
\renewcommand\nextpg{\pageref{pgs1}}\renewcommand\titleA{
The Role of Bag Surface Tension in Color Confinement 
}\renewcommand\authorA{
K. A. Bugaev${}^{\mathrm{a}}$,
 A. I. Ivanytskyi${}^{\mathrm{b}}$
}\renewcommand\email{
 e-mail:\space \eml{a}{bugaev@th.physik.uni-frankfurt.de},
 \eml{b}{a\underline{~}iv\underline{~}@ukr.net}\\[1mm]
}\renewcommand\titleH{
Role of Bag Surface Tension in Color Confinement 
}\renewcommand\authorH{
Bugaev K. A., Ivanytskyi A. I.
}\renewcommand\titleC{\titleA}\renewcommand\authorC{\authorH}\renewcommand\institution{
Bogolyubov Institute for Theoretical Physics, Kiev, Ukraine
}\renewcommand\abstractE{
We discuss here  the novel view at the color confinement which, 
on the one hand, allows us to find out the surface tension coefficient
of quark gluon bags and, under a plausible assumption, to determine the endpoint temperature of the QCD phase diagram, on the other hand. 
The present  model considers the confining color tube as the cylindrical 
quark gluon bag with non-zero surface tension. 
A close inspection of the free energies of elongated cylindrical  bag  and the confining  color  tube  that  connects the static quark-antiquark pair 
allows us to find out the string tension in terms of the surface tension, thermal pressure and the bag radius.   Using the derived relation it is possible
to estimate the bag surface tension at zero temperature directly from the lattice QCD data
and to estimate the (tri)critical endpoint temperature. 
In the present analysis the topological free energy of the cylindrical  bag is accounted for the first time.
The requirement of positive entropy density of such bags leads to 
negative values of the surface tension coefficient of quark gluon bags at the cross-over region, i.e. at the continuous transition to deconfined quarks and gluons.
We argue that the cross-over existence  at supercritical temperatures  in ordinary liquids is 
also provided by the negative surface tension coefficient values. 
It is shown that   the confining  tube model  naturally accounts for an existence of a very pronounced surprising  maximum  of the  tube   entropy observed  in the lattice QCD simulations, which, as we argue, signals about the fractional surface formation of
the confining  tube. In addition, using the developed  formalism we suggest  the gas of free tubes model and demonstrate that it contains  two  phases. }

\begin{article}

\section{Introduction}

A new paradigm of heavy ion phenomenology that the quark gluon plasma (QGP) is a strongly interacting liquid \cite{Bugaev:Ref1n} proved to be very successful not only in describing   some of its  properties measured by lattice quantum chromodynamics (QCD), but also in explaining  some experimental observables that cannot be reproduced otherwise.  
Probably, the two most striking conclusions obtained  within the  new paradigm are as follows:
first, at the cross-over temperature, where the  string tension of color tube is almost vanishing, the potential energy of color charge is of the order of a few GeV \cite{Bugaev:Ref2n}, i.e.  it is 10 times larger 
than its  kinetic energy, and, second,  the QGP, so far,  is the most perfect fluid since its shear viscosity in units of the entropy density is found to be the  smallest  one \cite{Bugaev:Ref3n,Bugaev:Ref4n}.
The first of these conclusions tells us that at the cross-over region there is no color charge separation \cite{Bugaev:Ref5n}, whereas the second one naturally  explains  the great success of ideal hydrodynamics when applied to relativistic heavy ion collisions.

Here we would like to discuss the recent progress achieved in our understanding of both the confinement phenomenon \cite{Bugaev:Ref6n,Bugaev:Ref7n} and the 
physical origin of the cross-over \cite{Bugaev:Ref8n,Bugaev:Ref9n,Bugaev:Ref10n,Bugaev:Ref11n,Bugaev:Ref12n}.  As we demonstrate below  such a progress was made possible after realizing a principal role played   by  negative values of the  surface tension coefficient of large QGP bags \cite{Bugaev:Ref6n,Bugaev:Ref8n}.  Also here we argue that the negative values of the  surface tension, that are responsible for an  existence of the cross-over transition to QGP at low baryonic densities, 
play the same role in ordinary liquids. 
Moreover, in this work we would like to draw an  attention to the problem of the  temperature dependence of surface tension coefficient in liquids by clearly showing  that for  many liquids  the well known  Guggenheim  relation (see Eq. (\ref{EqBugaevX}))  is not  so well established experimentally as it is usually believed. 

The work is organized as follows. Section 2 is devoted to the confining tube model, in which the Fisher topological term of the QGP bag free energy  is accounted for.
In section 3 we show that at the cross-over region the surface tension of QGP bags is necessarily negative and argue that this is the case for  ordinary liquids as well. 
The  maximum of the tube entropy observed in lattice QCD  is explained  in section 4,  where the model of gas of free tubes is also developed. The conclusions are given in the last section.

\section{Color confining tube and  sQGP}

A color confinement, i.e. an absence of  free color charges, is usually described  by  the free energy of heavy (static) quark-antiquark pair {\boldmath$F_{q\bar q} (T, L) = \sigma_{str} \cdot L$}. 
In the  lattice QCD  the functional dependence of 
$F_{q\bar q} (T, L)$ on the
temperature $T$ and the separation distance $L$ can be extracted  from the Polyakov line correlation  in a color singlet channel.  Then it is customary to define 
%

\begin{figure}[t]
\centering
\includegraphics[width=0.42\textwidth]{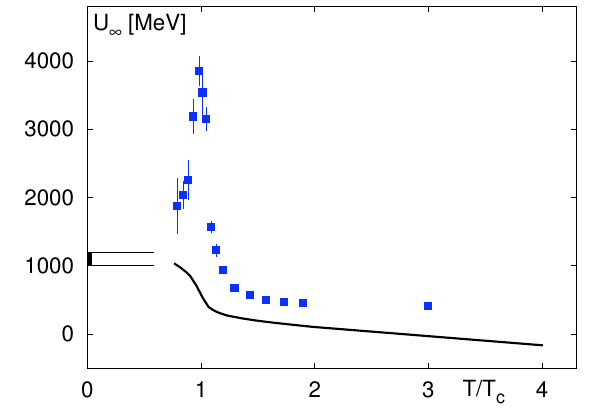}\hspace*{0.5cm}\includegraphics[width=0.42\textwidth]{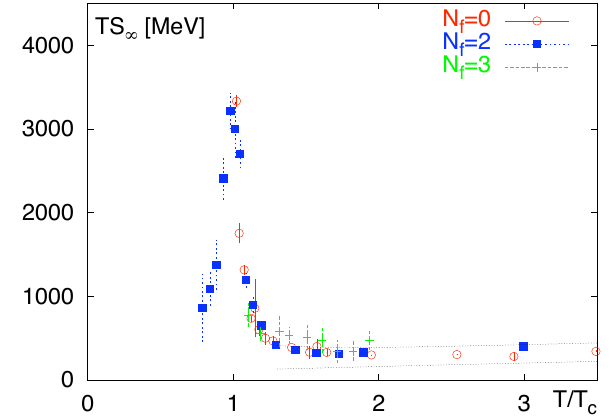}
  \caption{The internal energy $U_\infty$ (left) and entropy $S_\infty$ (right) of confining tube connecting two static color charges found by lattice  QCD simulations  for infinite separation distance between charges \cite{Bugaev:Ref2n}.
The internal energy is shown for 2 quark  flavors.}\label{Bugaev:fig:1}
\end{figure}

\begin{itemize}
%
\item {\bf confinement:} the case of non-zero string tension, i.e. {\boldmath$\sigma_{str} >0$};
%
\item {\bf deconfinement:} the case of vanishing  string tension {\boldmath$\sigma_{str} \rightarrow 0$}  at   $T \rightarrow T_{co}$,  but one should remember that   there is no color charge separation up to $T \ge 1.3 \,T_{co}$ values of  the cross-over temperature $T_{co}$ \cite{Bugaev:Ref1n,Bugaev:Ref5n}.
\end{itemize}
The explanation of the latter is as follows:  although at large distances $L \rightarrow \infty$  the potential energy of static $q\bar q$ pair   is finite $U_{q\bar q} (T, L) = F_{q\bar q} - T \frac{\partial F_{q\bar q}}{\partial T} = F_{q\bar q} + T S_{q\bar q}$ near $T_{co}$,  the values of $U (T, \infty)$ are very large (see Fig.~\ref{Bugaev:fig:1}). From Fig.~\ref{Bugaev:fig:1} one can conclude that 
near $T_{co}$ region QGP is a strongly interacting plasma  (sQGP) which is similar to a liquid,  since the ratio of the quark potential energy to its kinetic energy, the so called plasma parameter,  $\frac{U (T, \infty)}{3\,T} \in 1-10$ has the range of  values that is typical  for  ordinary  liquids \cite{Bugaev:Ref1n}.

The second striking feature of the confining tube can be seen  in the right panel of Fig.~\ref{Bugaev:fig:1} which clearly demonstrates that at $T=T_{co}$ the entropy of static $q\bar q$ pair  is very large 
$S_{q\bar q} (T_{co}, \infty) \approx 20$. Such a value  signals  that really a {\bf huge number of degrees of freedom $\sim \exp(20)$} is involved, but the origin of large energy $U_{q\bar q} (T, \infty)$ and entropy $S_{q\bar q} (T, \infty)$ values near $T_{co}$ for awhile  remained  {\bf mysterious} \cite{Bugaev:Ref1n} despite many attempts to explain it.

Another problem of  principal importance for phenomenological models of deconfinement phase transition \cite{Bugaev:Ref8n,Bugaev:Ref9n,Bugaev:Ref10n,Bugaev:Ref11n,Bugaev:Ref12n,Bugaev:Ref13o,Bugaev:Ref14o,Bugaev:Ref15o} is the value of  the surface tension coefficient $\sigma_{surf}$ of QGP bags. 
There are several estimates for the surface tension coefficient 
$\sigma_{surf}$ of QGP bags \cite{Bugaev:Ref13n},
but the question is whether  can we determine 
$\sigma_{surf}$ from lattice QCD?
Therefore, in this section we  consider an approach  that allows us to determine the 
surface tension coefficient  of QGP bags directly from the lattice QCD. As it will be shown  in the  section 4  such an approach naturally explains an existence of the `mysterious maximum' \cite{Bugaev:Ref1n} of  the confining tube entropy. 

In order to estimate the surface tension of QGP bags let us consider 
the static quark-antiquark pair connected by the unbreakable color tube of length $L$ and radius $R \ll L$.
In the limit of large $L$  the free energy of the color tube is $F_{str} \rightarrow \sigma_{str} L$. Now we  consider the same tube as an elongated cylinder of the same radius and length \cite{Bugaev:Ref6n}. 
In this case we neglect the free energy of the regions around the color charges, but for our treatment of large 
separation distances $L \gg R$ this is sufficient.
For the cylinder  free energy we use the standard  parameterization \cite{Bugaev:Ref8n,Bugaev:Ref9n,Bugaev:Ref10n,Bugaev:Ref11n,Bugaev:Ref12n}
\begin{eqnarray}\label{EqBugaevI}
&&\hspace*{-0.5cm}F_{cyl} (T, L, R)  = 
- p_v(T) \pi R^2 L + \sigma_{surf} (T) 2 \pi R L +  T \tau \ln\left[
\frac{\pi R^2 L }{V_0} \right] \,,
\end{eqnarray}
where $p_v (T)$ is the bulk pressure inside a bag, $\sigma_{surf} (T)$ is the temperature dependent 
surface tension coefficient,  while the 
last term on the right hand side above  is the Fisher topological term \cite{Bugaev:Ref14n}  which 
is proportional to the Fisher exponent $\tau = const > 1 
$ \cite{Bugaev:Ref8n,Bugaev:Ref9n,Bugaev:Ref10n,Bugaev:Ref11n,Bugaev:Ref12n,Bugaev:Ref15o} and $V_0 \approx 1$ fm$^3$ is a 
normalization constant.  
Since we consider the same object then its free energies calculated as the color tube  and as  the cylindrical bag should be equal to each other. Then for large separating distances $L \gg R$ one 
finds the following relation 
\begin{equation}\label{EqBugaevII}
\sigma_{str} (T) = \sigma_{surf} (T)\, 2 \pi R~ - ~p _v (T) \pi R^2 + \frac{T \tau}{L} \ln\left[
\frac{\pi R^2 L }{V_0} \right]  \,. 
%
\end{equation}
In doing so, in fact,  we match an ensemble of all string shapes of fixed $L$ to a mean elongated cylinder, 
which according to the original Fisher idea \cite{Bugaev:Ref14n} and the results   of the 
Hills and Dales Model (HDM) \cite{Bugaev:Ref15n,Bugaev:Ref16n} represents a sum of all surface deformations of such  a  bag.
The last equation allows one to determine the $T$-dependence of bag surface tension as
\begin{equation}\label{EqBugaevIII}
\sigma_{surf} (T) = \frac{\sigma_{str} (T)}{ 2 \pi R} ~ + ~ \frac{1}{2} \, p_v (T) R 
~ - ~ \frac{T \tau}{2 \pi R L} \ln\left[
\frac{\pi R^2 L }{V_0} \right] \,,
\end{equation}
if $R(T)$, $\sigma_{str} (T)$ and $p_v (T)$ are known. This relation opens a principal possibility to determine the bag surface tension  directly from the lattice QCD simulations for any $T$. 
Also it allows us to estimate  the surface tension at $T=0$. Thus, taking the typical value of the bag model pressure which is used in hadronic spectroscopy $p_v (T=0) = - (0.25)^4$ GeV$^4$ and inserting 
into (\ref{EqBugaevIII}) the lattice QCD values  $R=0.5$ fm  and $\sigma_{str} (T=0) = (0.42)^2$ GeV$^2$ 
\cite{Bugaev:Ref17n}, 
one finds  $\sigma_{surf} (T=0) = (0.2229~ {\rm GeV})^3 + 
0.5\, p_v\, R\approx (0.183~{\rm GeV})^3 \approx 157.4$ MeV fm$^{-2}$ \cite{Bugaev:Ref6n}.
The last term in (\ref{EqBugaevIII}) does not modify our above estimate at $T=0$, but, in contrast to \cite{Bugaev:Ref6n,Bugaev:Ref7n},  we keep it  in order to demonstrate its 
importance for the confining tube with free color charges.

The found value of the bag surface tension at zero temperature is very important for the phenomenological equations of state  of strongly interacting matter in two respects. 
Firstly, according to HDM the obtained value defines the temperature at which    the bag surface tension coefficient changes the sign \cite{Bugaev:Ref15n,Bugaev:Ref16n,Bugaev:Ref7n}
\begin{equation}\label{EqBugaevIV}
 T_\sigma = \sigma_{surf} (T=0)\, V_0^\frac{2}{3} \cdot \lambda^{-1}~ \in ~[148.4;~ 157.4] ~{\rm MeV} \, , 
\end{equation}
where the constant $\lambda = 1$ for the Fisher parameterization of the $T$-dependent  surface tension coefficient \cite{Bugaev:Ref14n} or 
$\lambda \approx 1.06009$, if we use the parameterization derived within the HDM for surface deformations \cite{Bugaev:Ref15n,Bugaev:Ref16n,Bugaev:Ref7n}.  
Secondly, according to one of the most successful  models of liquid-gas phase transition, i.e. the Fisher droplet model (FDM)  \cite{Bugaev:Ref14n}   the surface tension coefficient linearly depends on temperature. This conclusion is well supported by HDM and by microscopic models of vapor-liquid interfaces  \cite{Bugaev:Ref21n}.
Therefore,   the temperature $T_\sigma$ in (\ref{EqBugaevIV}), at which the surface tension coefficient vanishes,  is also  the  temperature of the (tri)critical endpoint 
$T_{cep}$ of the liquid-gas phase diagram.  On the basis of these arguments in Ref. \cite{Bugaev:Ref7n}  we concluded that  the value of QCD critical endpoint temperature is 
$T_{cep}= T_\sigma =  152.9 \pm 4.5 $ MeV.  Hopefully, the latter can be verified by the lattice QCD simulations using Eq. (\ref{EqBugaevIII}). 

Now the question is what is the surface tension coefficient above $T_{cep}$, i.e. at  supercritical temperatures. There are no experimental data on usual liquids in this region. 
In FDM and in the other well known model of liquid-gas phase transition, the
statistical multifragmentation model  (SMM) 
\cite{Bugaev:Ref18n,Bugaev:Ref19n,Bugaev:Ref20n},  the surface tension at supercritical temperatures is assumed to be zero, while in other models such a  question is usually not discussed.  The only exceptions known  to us are the exactly solvable statistical models of quark gluon bags with surface tension \cite{Bugaev:Ref8n,Bugaev:Ref9n},  their extension which includes the finite widths of large/heavy QGP bags 
\cite{Bugaev:Ref10n,Bugaev:Ref11n,Bugaev:Ref12n} and recently formulated  generalization of the SMM \cite{Bugaev:Ref125n}. For all  these models  it was demonstrated that the negative surface tension is the only physical reason of degeneration of the 1-st order phase transition into cross-over at supercritical temperatures. 
The question is whether the above suggested formalism can support such a conclusion.

\section{Surface tension coefficient  at the cross-over temperature}

The above results, indeed,  allow us to tune the interrelation with the color tube  model and to study the bag 
surface tension near the cross-over to QGP.  Consider first the vanishing baryonic densities. The lattice QCD data indicate that at large $R$ the 
string tension behaves as 
\begin{equation}\label{EqBugaevV}
 \sigma_{str} ~ =~ \frac{\ln\left( L/L_0 \right)}{R^2 } g_0 \,,
\end{equation}
where $L_0 > 0 $ and $g_0 > 0$ are some positive constants.  Assuming the validity of  Eq. (\ref{EqBugaevV}) 
in the infinite available 
volume, one finds that for  $\sigma_{str} (T)\rightarrow + 0$  the string radius diverges, i.e.
$R \rightarrow \infty$.

Using Eqs. (\ref{EqBugaevI}) and (\ref{EqBugaevIII}) we can write  the total pressure $p_{tot}$  of the cylinder as follows
\begin{eqnarray}
p_{tot} (L, R, T)& =& p_v (T)   - \frac{ \sigma_{surf} (T)}{R} - \frac{T \tau}{\pi R^2 L}  \equiv    \frac{ \sigma_{surf} (T)}{R} - \frac{\sigma_{str}}{\pi R^2}  +  \frac{T \tau}{\pi R^2 L}  \left[ \ln \left( 
\frac{\pi R^2 L}{V_0}  \right) - 1 \right] \nonumber \\
\label{EqBugaevVI}
& = &  \frac{ \sigma_{surf} (T)}{R} - \frac{g_0 \ln\left( L/L_0 \right)}{\pi R^4}  +  \frac{T \tau}{\pi R^2 L}  \left[ \ln \left( 
\frac{\pi R^2 L}{V_0}  \right) - 1 \right] \, .
\end{eqnarray}
This equation shows that for fixed separation distance $L$ in the limit $\sigma_{str} (T)\rightarrow + 0$  the leading term is given by the surface tension contribution, while 
the next to leading term corresponds to the contribution of the Fisher topological term, whereas the second term on the right hand side of  (\ref{EqBugaevVI}) is the smallest one. 
Therefore, it is evident that for small values of string tension (and large $R$) the main contribution to the total  pressure and to its temperature derivative is given by the first term 
on the right hand side of  (\ref{EqBugaevVI}).

To calculate the  total entropy density $s_{tot}$  of the cylinder let us  parameterize 
the string tension as
\begin{equation}\label{EqBugaevVII}
\sigma_{str}= \sigma_{str}^0\,  t^\nu
\end{equation}
where  $t \equiv \frac{T_{co} -T}{T_{co}} \rightarrow +0$   and $\nu =const >0$. 
 From (\ref{EqBugaevVII})  it follows $R = \left[ \frac{ g_0  \ln(L/L_0)}{ \sigma_{str}^0 t^\nu} \right]^\frac{1}{2}$ and then for $t \rightarrow ~0$  the  entropy density $s_{tot}$ can be found from (\ref{EqBugaevVI}) and (\ref{EqBugaevVII}) as
\begin{eqnarray}\label{EqBugaevVIII}
s_{tot} & = & \left(\frac{\partial ~p_{tot}}{\partial~T} \right)_\mu   \rightarrow  
\underbrace{- \frac{\nu}{2 \,R \, T_{co}}   \,  \frac{\sigma_{surf}}{t} }_{dominant~~since~~ 
t \rightarrow ~0 } \,+\,
 \frac{1}{R}  
\frac{\partial~\sigma_{surf} }{\partial ~T} \rightarrow   - ~  \frac{\nu}{2  \, T_{co}}  \left[ 
\frac{\sigma^0_{str}}{ g_0  \ln\left( L/L_0 \right)} \right]^\frac{1}{2} \,  \frac{\sigma_{surf}}{t^{1 - \nu/2}}
 ~> ~0 \, .
\end{eqnarray}
This equation shows that  at $T= T_{co}$ the entropy density diverges for $\nu < 2$ and also that  at the cross-over  region  the surface tension coefficient must be negative otherwise the system would be thermodynamically unstable since its entropy density would be negative.

\begin{figure}[ht]
\centering
  \includegraphics[width=0.84\textwidth]{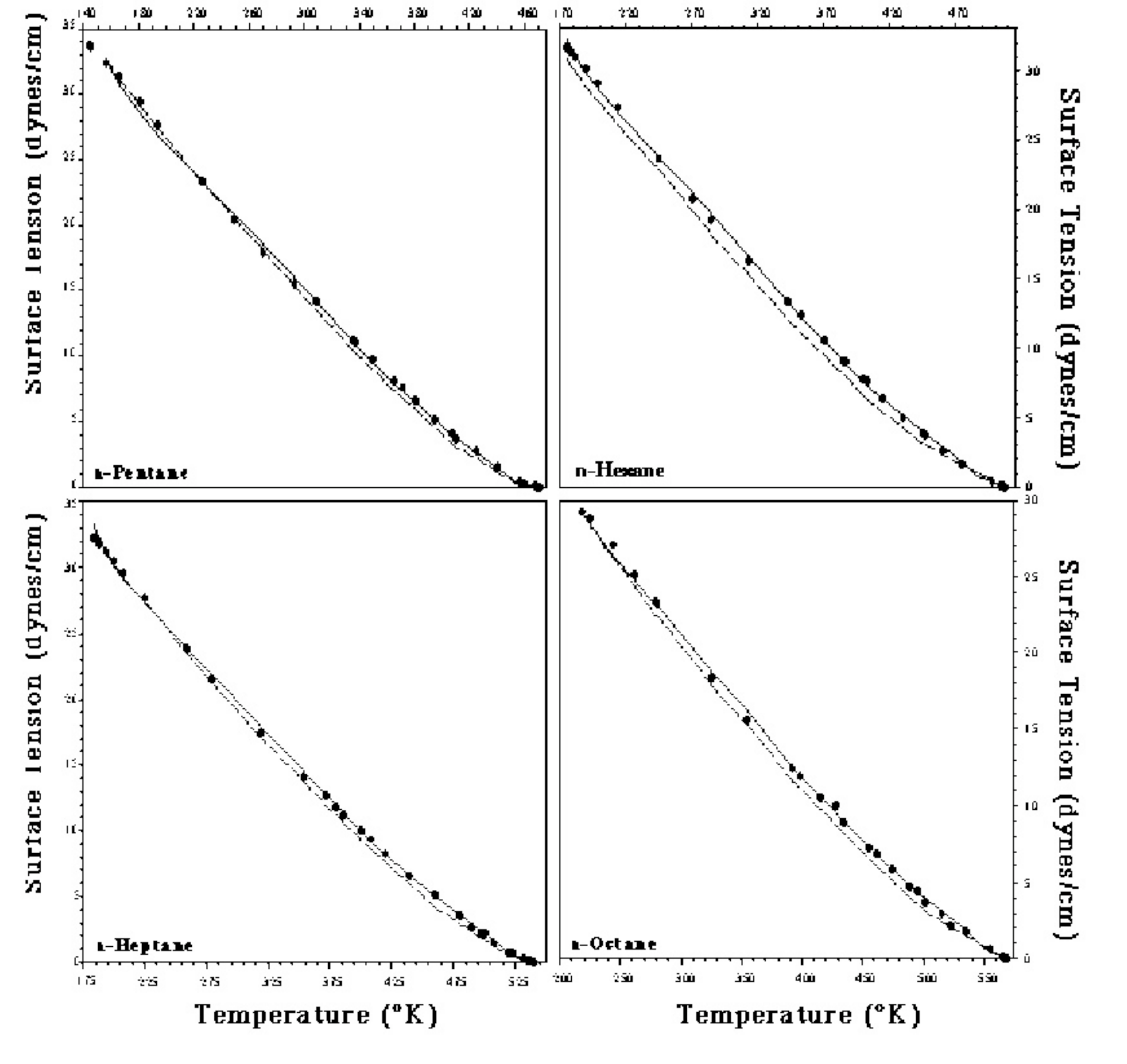}\\
  \caption{Surface tension  of normal paraffins as a function of temperature from the triple point to the 
critical point.  The filled circles indicate the experimental data \cite{Bugaev:Ref29n}.
The lines are the different theoretical  parameterizations \cite{Bugaev:Ref27n}.}\label{Bugaev:fig:2}
\end{figure}

Clearly, the results (\ref{EqBugaevI})--(\ref{EqBugaevVIII}) are valid for nonzero baryonic chemical potential $\mu$ up to the (tri)critical endpoint.  The main modification in (\ref{EqBugaevI})--(\ref{EqBugaevVIII})  is an appearance of $\mu$-dependences of $p_v (T, \mu)$ and $T_{co} (\mu)$ \cite{Bugaev:Ref6n}. 
In the (tri)critical endpoint vicinity
the  behavior  of   $p_{tot}$ and $s_{tot}$   is defined by the $T$-dependence of the  surface tension coefficient.

 We stress that there is nothing wrong 
or unphysical with the negative values of surface tension coefficient, since 
$ \sigma_{surf}\, 2 \pi  R L$ in (\ref{EqBugaevI}) is {\bf the surface  free energy} and, hence, as any free energy,  it contains the energy part
$E_{surf}$ and  the entropy part $S_{surf}$ multiplied by temperature $T$, i.e. $F_{surf}= 
E_{surf} - T S_{surf} $ \cite{Bugaev:Ref15n,Bugaev:Ref16n}. Therefore, at low temperatures the energy part dominates and the  surface  free energy is positive, whereas at high temperatures the number of   configurations  of a cylinder  with large surface  drastically increases  and  the surface free energy becomes negative since $S_{surf} > \frac{E_{surf}}{T}$.
Moreover, the exactly solvable models with phase transition and cross-over 
\cite{Bugaev:Ref8n,Bugaev:Ref9n,Bugaev:Ref10n} have region of  negative surface tension coefficient and they clearly show that the only reason why the 1-st order deconfinement phase transition degenerates into a cross-over at low baryonic densities is the negative values of $ \sigma_{surf}$ at this region and the above results independently prove this fact.

We believe that the same is true for many ordinary liquids otherwise one has to search for an alternative explanation for the disappearance of the 1-st order liquid-gas phase transition at the supercritical temperatures.  Of course, the experimental data in this region do not exists, but, nevertheless, there is indirect evidence for an existence of negative values of the surface tension coefficient at  the supercritical temperatures. To demonstrate the validity of this statement  we have to remind that 
the modern experimental data on the temperature dependence of the surface tension do not allow one to definitely conclude what is $T$-dependence at the vicinity of critical temperature $T_c$.
In fact there are two alternative prescriptions \cite{Bugaev:Ref26n,Bugaev:Ref26b}
\begin{eqnarray}\label{EqBugaevIX}
{\rm E\ddot{o}tv\ddot{o}s~~rule:} \quad  \quad  \frac{ \sigma_{surf}}{\rho_l^\frac{2}{3}} &=& a_E (T_c - T) \, , \\
{\rm Guggenheim~~rule:} \quad  \quad  \frac{ \sigma_{surf}}{\rho_l^\frac{2}{3}} &=& a_G (T_c - T)^n\, \quad {\rm with} \quad n \approx \frac{11}{9}\, ,
\label{EqBugaevX}
\end{eqnarray}
where $\rho_l$ is the temperature dependent particle density of the liquid phase.

\begin{figure}[ht]
\centering
  \includegraphics[width=1.0\textwidth]{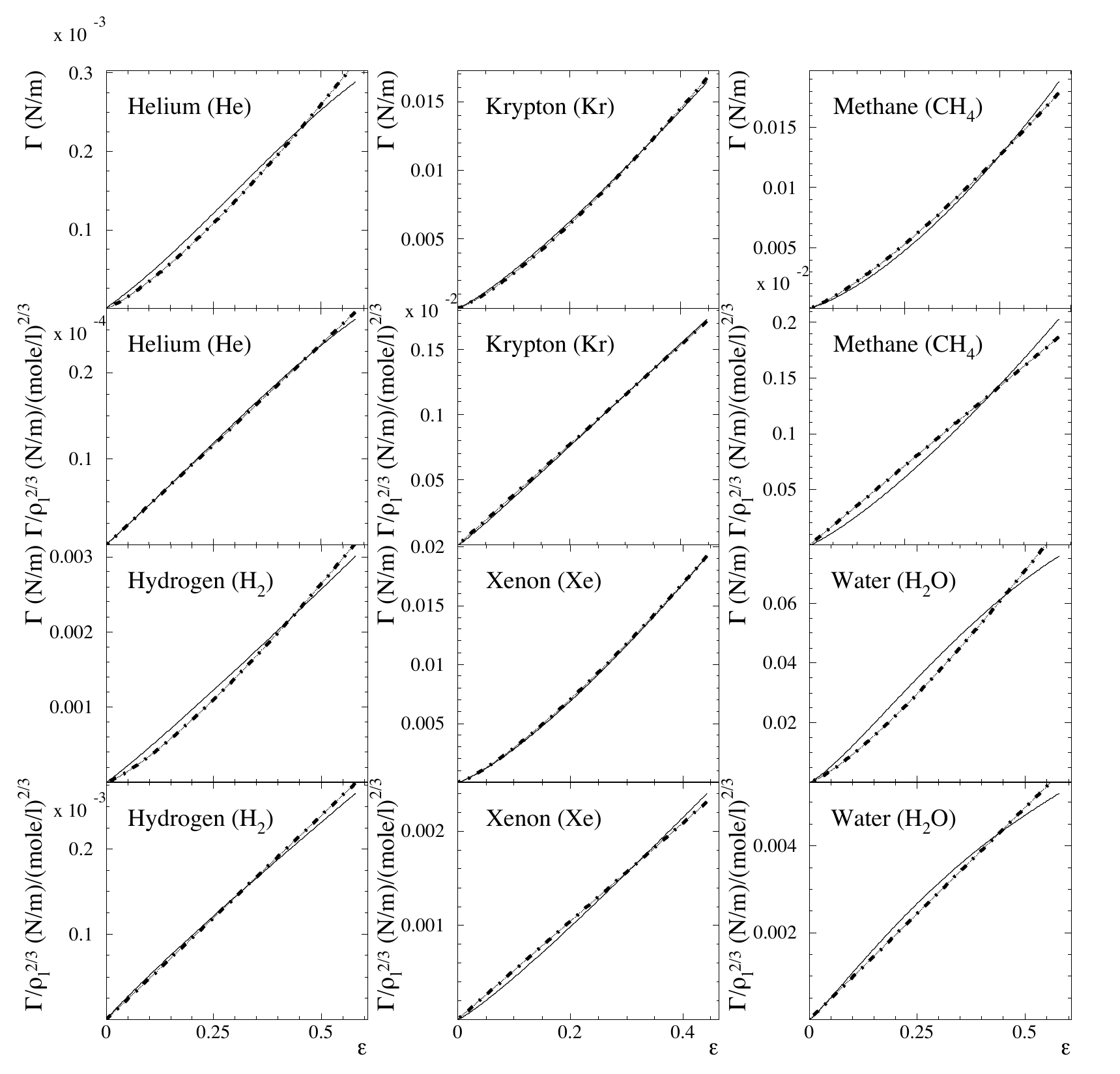}\\
  \caption{
The surface tension $\Gamma$ (in $N/m$) in terms of cluster surfaces (first and third rows) and the surface tension $\Gamma / \rho_l^{2/3}$ (in 
$(N/m)/(mole/l)^{2/3}$) 
in terms of cluster number (second and fourth rows)) as a function of $\varepsilon =(T_c-T)/T_c$ for: ÒquantumÓ fluids 
(hydrogen and helium), noble gases (krypton and xenon) and more complex fluids (methane and water). The thin solid lines show data points \cite{Bugaev:Ref33n} and the heavy dashed-dotted lines over the thin line show fits to  
$\Gamma = \Gamma_0\,  \rho_l^{2/3}  \varepsilon^{2 n} $ and $\Gamma / \rho_l^{2/3} = \Gamma_0\, \varepsilon$ according to 
Eqs. (\ref{EqBugaevX})  and  (\ref{EqBugaevIX}), respectively.
}
\label{Bugaev:fig:3}
\end{figure}


After  the Guggenheim work \cite{Bugaev:Ref26n} the prescription (\ref{EqBugaevX}) became
a dominant one \cite{Bugaev:Ref27n}. Sometimes there appeared even confusions.
Thus, in \cite{Bugaev:Ref28n} the authors determined the surface tension of water from the
triple point to critical  point and parametrized it by  the polynomial of 9-th power $ \sigma_{surf} (T) = \sum\limits_{l=1}^9 a_l \, (T_c - T)^l$, but then the same authors refitted it to the prescription 
(\ref{EqBugaevX}) \cite{Bugaev:Ref29n}. Here in Fig.~\ref{Bugaev:fig:2} which is taken from \cite{Bugaev:Ref27n} we show the temperature dependence 
of some paraffins.  As one can see for n-Pentane and n-Heptane the  data on temperature dependence of surface tension near the critical point, indeed, may show the nonlinear behavior similar to (\ref{EqBugaevX}), but for the n-Hexane and n-Octane one can see the linear $T$-dependence of Eq. (\ref{EqBugaevIX})!

Therefore, in order to clarify this issue a few years ago a thorough analysis  
\cite{Bugaev:Ref32n} of the high quality  NIST data  \cite{Bugaev:Ref33n} was performed.
Some of the results are shown in Fig.~\ref{Bugaev:fig:3} which is taken from Ref. \cite{Bugaev:Ref32n}. As one can see from Fig.~\ref{Bugaev:fig:3} for most of the analyzed liquids the linear prescription (\ref{EqBugaevIX}) provides an essentially better fit with the only exceptions of xenon and methane. 
Therefore, our first  conclusion is that for many liquids the rule (\ref{EqBugaevIX})  better describes the data than the rule (\ref{EqBugaevX}). 
The second conclusion one can draw from this discussion is that naive extrapolation of the linear $T$-dependence (\ref{EqBugaevIX}) of the surface tension coefficient $\sigma_{surf}$ to supercritical temperatures $T > T_c$ would lead to the negative values of the surface tension coefficient.
Of course, one may think that  $\sigma_{surf}  \equiv 0$ for  $T > T_c$ like in the FDM \cite{Bugaev:Ref14n} and SMM \cite{Bugaev:Ref18n,Bugaev:Ref19n,Bugaev:Ref20n}, but in this case  
one has to explain the reason why the $T$-derivative of $\sigma_{surf}$ has a discontinuity   
at $T = T_c$ while  the pressure and all its first and second derivatives are continuous functions of it arguments in this region. 

\section{The mysterious maximum of the lattice entropy and the gas of free tubes}

The considered configuration of the unbroken confining tube is only one of many other configurations accounted by the lattice QCD thermodynamics. However,
in order to explain a mysterious maximum of the lattice entropy (see Fig~\ref{Bugaev:fig:1}) it is sufficient  to  assume that the probability
of the unbroken confining tube among other configurations measured by lattice QCD is 
$W (L) \sim [L \, g_0 \ln (L/L_0)]^{-1}$, i.e. in the limit $L \rightarrow \infty$ it is negligible for any $\nu \neq 0$. Then the contribution of the unbroken confining tube into the lattice free energy is small,  since $W (L) F_{str} (L)  \sim R^{-2}  $ for $t \rightarrow +0$ and $R \rightarrow R_{lat}-0$ ($R_{lat}$ denotes the lattice size), but its contribution to 
the tube entropy 
\begin{equation}\label{EqBugaevXI}
W(L) S_{str} = - W \frac{d F_{str}}{d T} = W L \frac{\sigma_{str}^0 \nu}{T_{co}} t^{\nu-1} \rightarrow  W L \frac{ \nu}{T_{co}} \left[ \frac{ \sigma_{str}^0}{[g_0 \ln (L/L_0)]^{1- \nu}} \right]^\frac{1}{\nu} R^\frac{2(1-\nu)}{\nu} \sim R^\frac{2(1-\nu)}{\nu} 
\end{equation}
is an increasing function of the tube radius $R$ for $\nu <1$.  Clearly, if the available size of the lattice $R_{lat}$ would be infinite then the contribution of the unbroken tube would  diverge, but for finite lattice size one should observe a fast increase at $T \rightarrow T_{co}$. 

The physical  origin of  a  singular behavior of  the tube entropy  (\ref{EqBugaevXI}) encoded in $\nu < 1$  is  rooted in the formation of fractal surfaces of the confining tube in the cross-over temperature vicinity \cite{Bugaev:Ref6n}.  This can  be clearly  seen from the  power $\frac{2 (1-\nu)}{\nu}$ of  $R$ on the right hand side of (\ref{EqBugaevXI}) which is fractal for any $\nu \neq \frac{2}{2+n}  $ where  $n =  1, 2, 3, ...$ Moreover, the appearance of fractal structures at $T=T_{co}$ can be easily understood within 
our model, if one recalls that only at this temperature the fractal  surfaces  can emerge
at almost no energy costs due to almost zero total pressure (\ref{EqBugaevVII}).
An explanation of the tube entropy decrease for $t < 0$ is similar \cite{Bugaev:Ref6n,Bugaev:Ref7n}. It means that the fractal surfaces gradually disappear since for $T > T_{co}$ the tube gradually occupies  the whole available lattice volume.  

Here we also would like to consider a toy model based on the total pressure (\ref{EqBugaevVI}) of the confining tube,  but for 
the non-static (or free) quark-antiquark pair. In this case the parameter $L$ should be considered as a free parameter which has to be determined from the maximum of the total pressure (\ref{EqBugaevVI}). Finding from this condition the radius dependent separation distance $L_w (R)$ which corresponds to the most probable and the stable state of the free  confining tube one has to substitute it into expression for the pressure (\ref{EqBugaevVI}) and find the corresponding radius of the tube from the equation $p=p_{tot}(L_w (R), R, T)$  for the set of  given external pressure $p$ and  temperature $T$. Clearly the determination of  an extremum of the total pressure (\ref{EqBugaevVI}) with respect to $R$ with  subsequent finding  of the separation distance $L$ should give the same result, but the first way is technically easier. 

Instead of the quantitative analysis of the resulting pressure  $p_{tot}(L_w (R), R, T)$ which requires the knowledge of values of  all constants, i.e. $L_0$, $g_0$, $\tau$ e.t.c., we prefer to  discuss some qualitative properties of the model and show that it has two phases. Also we have to  stress that such a model cannot be applied  to high pressures (or high densities) directly because in this case the pressure of the system should account for the short range repulsion between  the tubes. Therefore in what follows it is assumed that the gas of tubes has some low particle  density $\rho$ and, hence,  one can  consider this gas as an ideal gas with the pressure $p = T \rho$. 

To determine  the density $\rho$ one has to maximize the pressure $p_{tot}$  first.
From the vanishing derivative condition
\begin{eqnarray}\label{EqBugaevXII}
\frac{\delta ~p_{tot} (L, R, T)}{\delta~L}  & = &  - \frac{g_0 }{\pi R^4 L}  - \frac{T \tau}{\pi R^2 L^2}  \left[ \ln \left( 
\frac{\pi R^2 L}{V_0}  \right) - 1 \right] + \frac{T \tau}{\pi R^2 L^2}  = 0 \, 
\end{eqnarray}
one can  find the following equation for $L_w(R)$
\begin{eqnarray}\label{EqBugaevXIII}
L_w & = &   \frac{T \tau R^2}{g_0 }    \left[2 -  \ln \left( 
\frac{\pi R^2 L_w}{V_0}  \right)  \right]  \, 
\end{eqnarray}
and show that it corresponds to a maximum of pressure. From Eq. (\ref{EqBugaevXII}) it is clearly seen  that, in contrast to previous findings \cite{Bugaev:Ref6n,Bugaev:Ref7n},  the role of the Fisher topological term is a decisive one  for an establishing Eq. (\ref{EqBugaevXIII}).

Consider a few limiting cases of Eq. (\ref{EqBugaevXIII}). In the limit $T\rightarrow 0$ and finite $R$ one gets  
\begin{equation}\label{EqBugaevXIV}
L_w^0 ~\approx ~ -   \frac{T \tau R^2}{g_0 }  \ln \left(  \frac{\pi R^4 \tau T}{V_0\, g_0 \, e^2}  \right) \rightarrow~0  \,,
\end{equation}
which can be interpreted as a confinement of color charges. 
The same solution is true for the case of $R \rightarrow 0$ and finite $T$. 
In the other extreme $T \rightarrow \infty$ (or  $R \rightarrow \infty$) one  obtains a different solution of Eq.  (\ref{EqBugaevXIII})
\begin{equation}\label{EqBugaevXV}
L_w^\infty~ \approx ~\frac{ V_0 e^2}{\pi R^2}   \,,
\end{equation}
which again shows that for large  $R$ values the separation distance is small. 
This situation resembles what is observed in lattice QCD for the non-static color charges:   the long tubes that connect  such charges  simply break up at some separation distances   \cite{Bugaev:Ref1n}. 

From Eq. (\ref{EqBugaevXIII}) one can show that the solution $L_w$ is a monotonically increasing function of $T$, while for $T \neq 0$  it always has a maximum as the function of $R$. 
Searching for the maximum  of $L_w(R)$  (\ref{EqBugaevXIII}) one can find the corresponding value of the radius $R_{max}$ and $L_w (R_{max})$
\begin{eqnarray}\label{EqBugaevXVI}
R_{max} & = &   \left[ \frac{g_0 \, V_0 \, e}{\pi \, \tau\,T}  \right]^\frac{1}{4}  \, ,\\
\label{EqBugaevXVII}
L_w (R_{max}) & = &   \left[  \frac{ V_0 \, e \, \tau \,T }{\pi \, g_0 }    \right]^\frac{1}{2}  \, ,
\end{eqnarray}
which, evidently,  obey the condition $\pi R_{max}^2 L_w (R_{max}) = V_0\, e $.
The  presence  of the maximum of function $L_w (R)$ leads to an existence of two different 
radii for the same value of separation distance $L$, or in other words, there are two solutions of the equation $L = L_w(R)$. Clearly, the parameters of the maximum $R_{max}$ and $L_w (R_{max})$ given, respectively,  by Eqs. (\ref{EqBugaevXVI}) and (\ref{EqBugaevXVII}), separate the regions of these solutions. 
Evidently,  the latter  correspond to two phases of the gas of tubes which have different tube
radius for the same separation distance L.
The analysis of these solutions shows that there are many possibilities which strongly depend on the values of the  involved constants $L_0$, $g_0$, $\tau$ and $V_0$, 
whereas for low  but non-vanishing temperatures  one can show that  the higher pressure corresponds to the phase of the tubes with smaller radius. This is clear because in case of  low temperatures  the leading contribution to  the total pressure (\ref{EqBugaevVI})  is  given by 
 the surface tension term $\frac{\sigma_{surf}}{R}$.

\section{Conclusions}  

In this work  we discuss the most general  relation between the tension of the color tube connecting the static quark-antiquark pair and the surface tension of the corresponding  cylindrical bag. Such a relation allows us to determine the surface tension of the QGP bags at zero temperature and,  under the plausible assumptions that are typical for  ordinary liquids,
 to estimate  the temperature of vanishing surface tension coefficient of QGP bags at zero baryonic charge  density as  $T_\sigma  =  152.9 \pm 4.5 $ MeV.
Using the Fisher conjecture  \cite{Bugaev:Ref14n} and  the exact results found for  the temperature dependence of surface tension coefficient  from   the partition of  surface deformations \cite{Bugaev:Ref15n,Bugaev:Ref16n,Bugaev:Ref7n},     we conclude that the same temperature range 
corresponds to the  value of QCD (tri)critical endpoint temperature, i.e. $T_{cep}= T_\sigma =  152.9 \pm 4.5 $ MeV.  
Then requiring the positive values for  the confining tube entropy density we demonstrate 
that at the cross-over region the surface tension coefficient of the QGP bags is unavoidably 
negative. 
Furthermore, analyzing the data on the temperature dependence of the surface tension coefficient of some ordinary liquids in the vicinity of the critical endpoint we conclude  that the negative values of the  surface 
tension coefficient of QGP bags   are not unique, but they  also  should exist at the supercritical temperatures of usual liquids. We believe such a conclusion is worth to verify experimentally for ordinary liquids. 

Also we demonstrate that the long unbroken tube taken with a vanishing probability 
which generates a finite contribution into the lattice free energy may, under certain assumptions, provide a very fast increase of the lattice entropy of such configurations
and, thus, it  may explain the  maximum  of the tube entropy observed by lattice QCD. 
Additionally, we considered    the non-static (free)  tube  and  used the developed formalism to work out the model of the gas of free tubes. The performed analysis of such a model showed that there are two phases in this model which correspond to different radii of  the tube for the same separation length $L$. 
Since this toy model resembles some important  features of the confining tubes observed in the lattice QCD, we believe it is important for QCD phenomenology and can be used to build up more  elaborate statistical  model which accounts for a realistic interaction between tubes.


\acknowledge{Acknowledgements}{
The research made in this work  
was supported in part   by the Program ``Fundamental Properties of Physical Systems 
under Extreme Conditions''  of the Bureau of the Section of Physics and Astronomy  of
the National Academy of Sciences  of Ukraine.
}


\end{article}

\label{pgs1}
\end{document}